\documentclass[12pt]{article}
\usepackage{amsfonts}
\usepackage{amsmath}
\usepackage{amssymb}
\usepackage{amscd}
\usepackage[dvips]{graphicx}

\input xy
\xyoption{all}

\begin{document}

\begin{center}
\textbf{\huge Braided algebras and the $\kappa $-deformed oscillators}

\vspace{0.8cm}

\textbf{Jerzy Lukierski and Mariusz Woronowicz}\\[0pt]
Institute for Theoretical Physics, University of Wroclaw\\[0pt]
pl. Maxa Borna 9, 50-206 Wroclaw, Poland\\[0pt]
lukier,woronow@ift.uni.wroc.pl

\vspace{1cm}

\textbf{Abstract}
\end{center}

Recently there were presented several proposals how to formulate the binary
relations describing $\kappa $-deformed oscillator algebras. In this paper
we shall consider multilinear products of $\kappa $-deformed oscillators
consistent with the axioms of braided algebras. In general case the braided
triple products are quasi-associative and satisfy the hexagon condition
depending on the coassociator $\Phi \in A\otimes A\otimes A$. We shall
consider only the products of $\kappa $-oscillators consistent with
co-associative braided algebra, with $\Phi =1$. We shall consider three
explicit examples of binary $\kappa $-deformed oscillator algebra relations
and describe briefly their multilinear coassociative extensions satisfying
the postulates of braided algebras. The third example, describing $\kappa $%
-deformed oscillators in group manifold approach to $\kappa $-deformed
fourmomenta, is a new result.

\section{Introduction.}

The notion of classical (pseudo)-Riemannian manifold as the geometric framework for the description of
gravitational effects in quantized theory appears to be not valid. The
dynamics of general relativity (Einstein equations) combined with
uncertainty principle of quantum mechanics imply that the exact
localization of space-time events is not possible. One can argue \cite{lit1}
that the quantum gravity effects cause the new algebraic uncertainty
relation between the relativistic space-time coordinates (usually we assume
that $\kappa $ is related with Planck mass $M_{p}$.)

\begin{equation}
\lbrack \hat{x}_{\mu }\,,\hat{x}_{\nu }]=\frac{1}{\kappa ^{2}}\Theta _{\mu
\nu }(\kappa \hat{x})\,\,,  \label{1g1}
\end{equation}%
where using Taylor expansion one gets

\begin{equation}
\Theta _{\mu \nu }(\kappa \hat{x})=\Theta _{\mu \nu }^{(0)}+\kappa \Theta
_{\mu \nu }^{(1)\rho }\hat{x}_{\rho }+\cdots \,.  \label{tt2}
\end{equation}

The first term in (\ref{tt2}) describes the canonical deformation, with
constant value of the commutator (\ref{1g1}), and second one introduces
so-called Lie-algebraic deformation of Minkowski space geometry. The general
formula (\ref{1g1}) has been also derived as the relation characterizing the $D$-brane
coordinates, which describe the end points of $D=10$ string \cite{lit4}.

An interesting task is to consider the construction of dynamic theories, in
particular QFT on the deformed Minkowski space (\ref{1g1}). In order to
formulate the deformed perturbation expansion of interacting quantized
fields it is important to study the theory of deformed free quantum fields.
In this paper we shall consider the case of so-called $\kappa $-deformation
\cite{kappa}-\cite{lrz} which is the well-known example of Lie-algebraic
deformation.

In order to deform consistently the relativistic free quantum fields one
should (see e.g.\cite{dlwd})

\begin{enumerate}
\item replace the space-time field arguments $x_{\mu }$ by their suitably
deformed noncommutative counterparts $\widehat{x}_{\mu }.$

\item in place of the classical field oscillators algebra $(p=(%
\overrightarrow{p},p_{0}=\omega (\overrightarrow{p})))$%
\begin{eqnarray}
&&[\;a(p),a(q)\;]=[\;a^{\dag }(p),a^{\dag }(q)\;]=0,  \label{ka1} \\
&&[\;a(p),a^{\dag }(q)\;]=\delta ^{(3)}(\vec{p}-\vec{q}),  \label{kappaccr2}
\end{eqnarray}%
introduce the deformed set of oscillator algebra relations.
\end{enumerate}

Dealing with noncommutative space-time arguments can be avoided if we can
represent homomorphically the algebra of noncommutative fields $\varphi (%
\widehat{x})$ by the suitable $\star $-product (see e.g.\cite{klm}-\cite{gf}%
) of standard commutative fields $\varphi (x)$
\begin{equation}
\varphi (\widehat{x})\chi (\widehat{x})~~\longleftrightarrow \;\varphi
(x)\star \chi (x)\;,\;\;~~\;\;\;\;~~\;  \label{3}
\end{equation}%
with the noncommutative structure contained in the proper choice of $\star $%
-product. It can be added that in order to describe e.g. in noncommutative
field theory the $n$-point Green functions we should extend the binary
representation (\ref{3}) to the $n$-ary products
\begin{equation}
\varphi _{1}(\widehat{x}_{1})\ldots \varphi _{n}(\widehat{x}%
_{n})~~\longleftrightarrow \;\varphi _{1}(x_{1})\star \ldots \star \varphi
_{n}(x_{n}),\;\;~~\;\;\;\;~~\;
\end{equation}%
where the $n$-tuple $(\widehat{x}_{1,\ldots ,}\widehat{x}_{n})$ provides the
extension of basic algebra describing simple noncommutative space-time point
to the algebraic variety of $n$ noncommutative space-time points.

The aim of this paper is to consider the deformations of the oscillator
algebras (\ref{ka1})-(\ref{kappaccr2}) obtained in $\kappa $-deformed
relativistic framework \cite{kappa}-\cite{lrz} with the noncommutative $%
\kappa $-Minkowski space \cite{zak}, \cite{mr}, \cite{lrz}
\begin{equation}
\lbrack \;{\hat{x}}_{0},{\hat{x}}_{i}\;]=\frac{i}{\kappa }{\hat{x}}%
_{i}\;\;,\;\;[\;{\hat{x}}_{i},{\hat{x}}_{j}\;]=0\;.  \label{kom}
\end{equation}%
Formula (\ref{kom}) is the special case of generalized $\kappa $-deformed
relation \cite{kmls}-\cite{dgp}
\begin{equation}
\lbrack \;{\hat{x}}_{\mu },{\hat{x}}_{\nu }\;]=\frac{i}{\kappa }(a_{\mu }{%
\hat{x}}_{\nu }-a_{\nu }{\hat{x}}_{\mu }),\qquad \;\;[\;{\hat{x}}_{i},{\hat{x%
}}_{j}\;]=0\;,  \label{gkom}
\end{equation}%
if we put $a_{\mu }=(1,0,0,0).$

The $\kappa $-deformed Poincare symmetries are described by dual pair of
Hopf algebras, the $\kappa $-deformed Poincare algebra and the $\kappa $%
-deformed Poincare group, with their explicit form depending on the choice
of basic generators. In this paper we shall use the following modified
Majid-Ruegg basis \cite{alz}, \cite{ga}, \cite{dlwd}

\begin{description}
\item[a)] algebraic sector $(\widehat{M}_{\mu \nu }=(\widehat{M}_{i}=\frac{1%
}{2}\epsilon _{ijk}\widehat{M}_{jk},\ \widehat{N}_{i}=\widehat{M}_{i0}))$
\end{description}

\begin{equation}
\lbrack \;\widehat{M}_{\mu \nu },\widehat{M}_{\lambda \sigma }\;]=i\left(
\eta _{\mu \sigma }\widehat{M}_{\nu \lambda }-\eta _{\nu \sigma }\widehat{M}%
_{\mu \lambda }+\eta _{\nu \lambda }\widehat{M}_{\mu \sigma }-\eta _{\mu
\lambda }\widehat{M}_{\nu \sigma }\right) \;,\;\;\;\;~~\;\;\;\;~~\;\;\;\;~~\;
\label{a1}
\end{equation}%
\begin{equation}
\lbrack \;\widehat{M}_{i},\widehat{P}_{j}\;]=i\epsilon _{ijk}\widehat{P}%
_{k}\;,\;\;\;\;\;\;\;\;\;~~\;\;\;\;~~\;\;\;\;~~\;\;\;\;\;\;\;~~\;\;\;\;~~\;%
\;\;\;~~\;\;\;\;\;\;~~\;\;\;\;~~\;\;\;\;~~\;\;  \label{a2}
\end{equation}%
\begin{equation}
\lbrack \;\widehat{N}_{i},\widehat{P}_{j}\;]=i\delta _{ij}\mathrm{e}^{\frac{%
\widehat{P}_{0}}{2\kappa }}\left[ \frac{\kappa }{2}\left( 1-\mathrm{e}^{-%
\frac{2\widehat{P}_{0}}{\kappa }}\right) +\frac{1}{2\kappa }\mathrm{e}^{-%
\frac{\widehat{P}_{0}}{\kappa }}\widehat{P}_{k}\widehat{P}_{k}\right] -\frac{%
i}{2\kappa }\mathrm{e}^{-\frac{\widehat{P}_{0}}{2\kappa }}\widehat{P}_{i}%
\widehat{P}_{j}\;,  \label{a3}
\end{equation}%
\begin{equation}
\lbrack \;\widehat{M}_{i},\widehat{P}_{0}\;]=0\;\;,\;\;\;\;[\;\widehat{N}%
_{i},\widehat{P}_{0}\;]=i\mathrm{e}^{-\frac{\widehat{P}_{0}}{2\kappa }}%
\widehat{P}_{i}\;,~~\;\;\;\;\;\;~~\;\;\;\;\;\;~~\;\;\;\;\;\;~~\;\;\;\;\;\;~~%
\;\;\;\;\;  \label{a4}
\end{equation}%
\begin{equation}
\lbrack \;\widehat{P}_{\mu },\widehat{P}_{\nu
}\;]=0\;,\;~~\;\;\;\;\;\;~~\;\;\;\;\;\;~~\;\;\;\;\;~~\;\;\;\;\;\;~~\;\;\;\;%
\;\;~~\;\;\;\;\;~~\;\;\;\;\;\;~~\;\;\;\;\;\;~~\;\;\;~~~~~  \label{a5}
\end{equation}

\begin{description}
\item[b)] coalgebra sector
\begin{equation}
\Delta (\widehat{M}_{i})=\widehat{M}_{i}\otimes 1+1\otimes \widehat{M}%
_{i}\;,\;\;\;\;~~\;\;\;\;\;\;\;~~\;\;\;\;\;\;\;~~\;\;\;\;\;\;\;~  \label{c1}
\end{equation}%
\begin{equation}
\Delta (\widehat{N}_{i})=\widehat{N}_{i}\otimes 1+\mathrm{e}^{-\frac{%
\widehat{P}_{0}}{\kappa }}\otimes \widehat{N}_{i}+\frac{1}{\kappa }\epsilon
_{ijk}\mathrm{e}^{-\frac{\widehat{P}_{0}}{2\kappa }}\widehat{P}_{j}\otimes
\widehat{M}_{k}\;,  \label{c2}
\end{equation}%
\begin{equation}
\Delta (\widehat{P}_{0})=\widehat{P}_{0}\otimes 1+1\otimes \widehat{P}%
_{0}\;,\;\;\;\;~~\;\;\;\;\;\;\;~~\;\;\;\;\;\;\;~~\;\;\;\;\;\;\;~~  \label{c3}
\end{equation}%
\begin{equation}
\Delta (\widehat{P}_{i})=\widehat{P}_{i}\otimes \mathrm{e}^{\frac{\widehat{P}%
_{0}}{2\kappa }}+\mathrm{e}^{-\frac{\widehat{P}_{0}}{2\kappa }}\otimes
\widehat{P}_{i}\;,\;\;\;\;~~\;\;\;\;\;\;\;~~\;\;\;\;\;\;\;~~\;  \label{c4}
\end{equation}

\item[c)] \bigskip antipodes
\begin{equation}
S(\widehat{M}_{i})=-\widehat{M}_{i}\;\;,\;\;S(\widehat{N}_{i})=-\mathrm{e}^{%
\frac{\widehat{P}_{0}}{\kappa }}\widehat{N}_{i}+\epsilon _{ijk}\mathrm{e}^{%
\frac{\widehat{P}_{0}}{2\kappa }}\widehat{P}_{j}\widehat{M}_{k}-\frac{1}{%
\kappa }\widehat{P}_{i}\mathrm{e}^{\frac{\widehat{P}_{0}}{2\kappa }}\;,
\label{anti1}
\end{equation}%
\begin{equation}
S(\widehat{P}_{i})=-\widehat{P}_{i}\;\;,\;\;S(\widehat{P}_{0})=-\widehat{P}%
_{0}\;.  \label{anti2}
\end{equation}%
One can show that the $\kappa $-deformed mass Casimir takes the form
\end{description}

\begin{equation}
C_{2}^{\kappa }(\widehat{P}_{\mu })=\left( 2\kappa \sinh \left( \frac{%
\widehat{P}_{0}}{2\kappa }\right) \right) ^{2}-\widehat{P}_{i}\widehat{P}%
_{i}=m^{2}\quad \Rightarrow \quad p_{0}=\omega _{\kappa }(\vec{p}%
^{2})=2\kappa ~\mathrm{arcsinh}\left( \frac{\sqrt{\vec{p}^{2}+M^{2}}}{%
2\kappa }\right) \;.  \label{mshell}
\end{equation}

We mention that the relations (\ref{a1})-(\ref{mshell}) can be obtained from
the ones in standard Majid-Ruegg (MR) basic \cite{mr} if we redefine the
three-momentum generators $\widehat{P}_{i}=e^{\widehat{P}_{0}/2\kappa }%
\widehat{P}_{i}^{MR}.$

The basic problem which should be addressed in the procedure of $\kappa $%
-deformation of relations (\ref{ka1})-(\ref{kappaccr2}) is the non-Abelian
structure of the $\kappa $-deformed addition law of momenta. It follows from
(\ref{c4}) that the two-particle $\kappa $-deformed momenta compose in the
following way:
\begin{equation}
\vec{p}^{(1+2)}=\vec{p}_{(1)}\mathrm{e}^{p_{0}^{(2)}/2\kappa }+\mathrm{e}%
^{^{-p_{0}^{(1)}/2\kappa }}\vec{p}_{(2)},\qquad
p_{0}^{(1+2)}=p_{0}^{(1)}+p_{0}^{(2)}\;.  \label{addd}
\end{equation}

The formulae (\ref{addd}) are derived from the Hopf-algebraic action of the
symmetry generators $\hat{g}$ on the products of representation modules a($%
\overrightarrow{p}$)
\begin{equation}
\hat{g}_{A}\triangleright \left( a(\vec{p})a(\vec{q})\right) =\left( \hat{g}%
_{A}^{(1)}\triangleright a(\vec{p})\right) \left( \hat{g}_{A}^{(2)}%
\triangleright a(\vec{q})\right) \;,\vspace{0.1cm}  \label{amod}
\end{equation}%
where $\Delta (\hat{g}_{A})=\hat{g}_{A}^{(1)}\otimes \hat{g}_{A}^{(2)}$ (we
put $\hat{g}_{A}$\bigskip $\equiv \hat{P}_{\mu }$, and $\widehat{P}_{\mu
}\triangleright a(\overrightarrow{p})=p_{\mu }a(\overrightarrow{p})(p_{\mu
}=(\vec{p},p_{0}))$.

We point out here that by introduction of $\kappa $-deformed oscillators we
shall obtain their $\kappa $-deformed algebra covariant under the
noncommutative translations, generated by $\widehat{P}_{\mu }$.

We shall consider here three ways of writing consistent $\kappa $%
-deformation of oscillator algebra (\ref{ka1})-(\ref{kappaccr2}).

\begin{enumerate}
\item By introducing new multiplication in the standard oscillator algebra (%
\ref{ka1})-(\ref{kappaccr2}) \cite{mpla}, \cite{dlwd}, \cite{jpa}.

\item By $\kappa $-deforming the flip operator $\tau _{0}:\tau _{0}[a(\vec{p}%
)a(\overrightarrow{q})]=a(\overrightarrow{q})a(\overrightarrow{p})$ \cite{am}%
- \cite{yz2}, \cite{jpa} in a way consistent with the addition law (\ref%
{addd}).

\item By replacing the composition of fourmomenta by suitable group addition
law \cite{glik}, \cite{glo} in consistency with the ideas of group field
theory \cite{fl}, \cite{oriti}.
\end{enumerate}

Our main aim here is to discuss the extension of these three sets of\ binary
relations to multilinear product of oscillators and show its consistency
with the axioms of braided algebras \cite{majid},\cite{found}. Firstly in
Sect. 2, we shall recall these axioms, in particular the hexagon relation
which is satisfied by triple braided products. We shall use only the hexagon
relation with coassociator $\Phi =1$, what leads to simpler version of
braided framework. In Sect. 3 we shall show explicitly how looks the
corresponding triple products for three mentioned above choices of $\kappa $%
-deformed algebras.

It should be mentioned that we shall not investigate here the problem of
covariance under full $\kappa $-deformed Poincare symmetries (see however
Sect. 4). It has been argued \cite{yz3}, \cite{yz4} that the requirement of
covariance under the $\kappa $-boosts leads necessarily to nontrivial choice
of $\Phi .$

\section{\protect\bigskip Braid factors and hexagon relations.}

In standard field theory in D=4 there are two statistics, bosonic and
fermionic, corresponding to the following two choices of tensor products
\begin{equation}
\Psi (a\otimes b)\equiv b\otimes a=\pm a\otimes b.
\end{equation}

For general braided algebra $A$ there exists an isomorphism $\Psi $ which in
general case does not satisfy the involutive condition $\Psi \circ \Psi =1.$
In such a case one can introduce second different braiding functor $\Psi
^{-1}$which leads in alternative way to the transposed product $b\otimes a.$
The restrictive choice provided if $\Psi $ is an involution
\begin{equation}
\Psi ^{-1}=\Psi \qquad \Longrightarrow \qquad \Psi \circ \Psi =1,
\end{equation}%
leads to simpler symmetric braided category. For tensor products $A^{\otimes
n}$ one can define the braid functor $\Psi _{(n,m)}$ transposing the tensor
products of algebra $A$
\begin{equation}
\Psi _{(n,m)}[(a_{1}\otimes \ldots \otimes a_{n})\otimes (b_{1}\otimes
\ldots \otimes b_{m})]=(b_{1}\otimes \ldots \otimes b_{m})\otimes
(a_{1}\otimes \ldots \otimes a_{n}),
\end{equation}%
and $\Psi _{(1,1)}\equiv \Psi .$

The triple products of general braided algebra can be transposed in two
equivalent ways. The transposition in general case require the coassociator
mapping $\Phi \in A\otimes A\otimes A$
\begin{equation}
\Phi ^{-1}(a\otimes (b\otimes c))\equiv (a\otimes b)\otimes c,
\end{equation}%
which should satisfy the pentagon relation \cite{found}. The binary braiding
factor $\Psi $ and coassociator $\Phi $ permits to relate the cyclic
transposition
\begin{equation}
a\otimes (b\otimes c)\quad \rightarrow \quad ((c\otimes a)\otimes b),
\end{equation}%
in accordance with the hexagon diagram \cite{found}
\begin{equation*}
\xymatrix@C=-25pt@!C{\ & a\otimes (b\otimes c) \ar[dr]^{\Phi^{-1}} \ar[dl]%
_{id\otimes \Psi} & & & & (a\otimes b)\otimes c \ar[dr]^{\Psi\otimes id} %
\ar[dl]_{\Phi}& \\
a\otimes (c\otimes b) \ar[d]_{\Phi^{-1}}& & (a\otimes b)\otimes c \ar[d]%
^{\Psi} & & a\otimes(b\otimes c) \ar[d]_{\Psi}& & (b\otimes a)\otimes c %
\ar[d]^{\Phi} \\
(a\otimes c)\otimes b \ar[dr]_{\Psi\otimes id}& & c\otimes(a\otimes b)\ar[dl]%
^{\Phi^{-1}} & & (b\otimes c)\otimes a \ar[dr]_{\Phi}& & b\otimes(a\otimes c)%
\ar[dl]^{id\otimes \Psi} \\
& (c\otimes a)\otimes b & & & &b\otimes(c\otimes a)\\
}
\end{equation*}

\begin{center}
\bigskip \textit{Fig. 1. Hexagon condition}
\end{center}

If $\Phi \equiv 1$ in the triple product we need not to use the subbrackets
\begin{equation}
a\otimes (b\otimes c)\equiv (a\otimes b)\otimes c=a\otimes b\otimes c,
\end{equation}%
and two hexagon relations degenerate into one triangle diagram
\begin{equation*}
\xymatrix@C=-10pt@W=1pt{&\ a\otimes b\otimes c\ar[dl]_{1\otimes \Psi }\ar[dr]%
^{\Psi _{(2,1)}}&\\
a\otimes c\otimes b\ar[rr]_{\Psi \otimes1} && c\otimes a\otimes b}
\end{equation*}

\begin{center}
\bigskip \textit{Fig. 2. Triangle diagram}
\end{center}

i.e. we see that
\begin{equation}
\Psi _{(2,1)}\equiv (\Psi \otimes 1)\circ (1\otimes \Psi ).  \label{brrr}
\end{equation}

In general case one can consider the braided structure of two different
braided algebras $A,B$ provided we known the braided factor $\widetilde{\Psi
}$ exchanging the elements $a\in A$ and $b\in B$.

In our discussion of $\kappa $-deformed oscillators we shall consider the
deformations of algebra (\ref{ka1},\ref{kappaccr2}) \ in the category of
associative braided algebras. Following (\ref{brrr}) , the braided factors $%
\Psi _{(1,2)},\Psi _{(2,1)}$ are the following products of binary braided
factors
\begin{equation}
\Psi _{(A,B\otimes C)}=\Psi _{(A,B)}\Psi _{(A,C)},\qquad \Psi _{(A\otimes
B,C)}=\Psi _{(A,C)}\Psi _{(B,C)}.
\end{equation}

It should be stressed that the quantum groups described by quasi-triangular
Hopf algebras define braided algebras with uniquely deformed braid factor $%
\Psi _{H}$ \cite{majid}, \cite{found}. The quasi-triangular Hopf algebra $%
(H,\Delta ,\epsilon ,S,R)$ is a Hopf algebra with universal $R$-matrix
determining quantum quasi-triangular structure relations $(\Delta \equiv
\Delta ^{(1)}\otimes \Delta ^{(2)};\Delta ^{op}\equiv \Delta ^{(2)}\otimes
\Delta ^{(1)})$
\begin{equation}
(1\otimes \Delta )R=R_{12}R_{13},\qquad (\Delta ^{op}\otimes
1)R=R_{23}R_{13},  \label{trih}
\end{equation}%
where $R$ is defined by the relation

\begin{equation}
\Delta ^{op}=R\Delta R^{-1},  \label{rr}
\end{equation}%
and $R_{12}=R\otimes 1$ etc. It appears that the category of $H$%
-representations $C=$Rep$(H)$ is a braided tensorial category with the braid
factor
\begin{equation}
\Psi _{H}(x\otimes y)=\tau (R\vartriangleright (x\otimes
y))=(R^{(1)}\vartriangleright y)(R^{(2)}\vartriangleright x),  \label{bra}
\end{equation}%
where $x,y\in $ Rep$(H)$ and $R=R^{(1)}\otimes R^{(2)}.$ The braiding (\ref%
{bra}) is covariant under the action of the Hopf algebra generators $h\in H$%
, i.e. one can show that
\begin{equation}
h\vartriangleright \Psi _{H}(x\otimes y)=\Psi _{H}[h\vartriangleright
(x\otimes y)],  \label{q}
\end{equation}%
in virtue of the relations (\ref{rr}). We see therefore that the covariant
representations of Hopf algebra $H$ ($H$-modules) are described by the
category of braided monoidal category of modules with braid factor $\Psi
_{H} $; usually such a module can be extended to a braided algebra.

From these general considerations follows immediately that for any twisted
quantum group $H_{F}$ it is possible to introduce the covariant braided
algebra of $H_{F}$-modules. The twist factor $F\in A\otimes A$ determines
the universal $R$-matrix by the formula \cite{yz3}-\cite{drin}
\begin{equation}
R=F^{T}F^{-1},  \label{w}
\end{equation}%
and from the relations (\ref{q})-(\ref{w}) we obtain the explicit formula
for the twist factor.

Unfortunately, in the case of $\kappa $-deformation of relativistic
symmetries the $\kappa -$deformed Poincare-Hopf algebra can not be obtained
by twisting procedure. At present there were proposed however several models
of $\kappa -$deformed oscillators, without the knowledge of universal $R$%
-matrix for $\kappa $-Poincare algebra. Our aim here is to show how these
various models are incorporated into the framework of braided algebras. In
our last Section 4 following \cite{yz1}, \cite{yz2}, \cite{yz3}, \cite{yz4}
we shall discuss briefly how one can look for the $\kappa $-deformation of
oscillators algebra (\ref{ka1})-(\ref{kappaccr2}), which is as well
covariant under the $\kappa $-Poincare transformations.

\section{\protect\bigskip Three examples of $\protect\kappa $-deformed
braided oscillators algebras}

\bigskip In ref. \cite{mpla} it was firstly observed that the consistency of
$\kappa $-deformed oscillator algebra with $\kappa $-deformed four-momentum
conservation law leads after the exchange of deformed oscillators to the
modification of the fourmomentum dependence. In such momentum-dependent
statistics, describing e.g. a pair of $\kappa $-deformed particles, the
behavior of one particle depends on the energy of the second one. One can
say that $\kappa $-deformation introduces some geometric interparticle
interaction modifying standard factorization properties of free particle
states.

In this Section we shall provide three versions of $\kappa $-deformed
oscillator algebra.\bigskip

Let us write down the general momentum-dependent $\kappa $-deformation of
the commuting creation oscillators as the following change of classical
exchange relations (see(\ref{ka1}))
\begin{equation}
a_{\kappa }(\overrightarrow{P},P_{0})a_{\kappa }(\overrightarrow{Q}%
,Q_{0})=a_{\kappa }(\overrightarrow{R},R_{0})a_{\kappa }(\overrightarrow{S}%
,S_{0}),  \label{ger}
\end{equation}%
where
\begin{eqnarray}
\overrightarrow{P} &=&\overrightarrow{P}(\overrightarrow{p},\overrightarrow{q%
}),\qquad \overrightarrow{Q}=\overrightarrow{Q}(\overrightarrow{p},%
\overrightarrow{q}),\qquad ~\overrightarrow{R}=\overrightarrow{R}(%
\overrightarrow{p},\overrightarrow{q}),\qquad \overrightarrow{S}=%
\overrightarrow{S}(\overrightarrow{p},\overrightarrow{q}),\qquad  \label{fo1}
\\
P_{0} &=&\omega _{\kappa }^{1}(\overrightarrow{p},\overrightarrow{q}),\qquad
Q_{0}=\omega _{\kappa }^{2}(\overrightarrow{p},\overrightarrow{q}),\qquad
R_{0}=\widetilde{\omega }_{\kappa }^{2}(\overrightarrow{p},\overrightarrow{q}%
),\qquad S_{0}=\widetilde{\omega }_{\kappa }^{1}(\overrightarrow{p},%
\overrightarrow{q}).  \label{fo2}
\end{eqnarray}%
We explicitly introduced for every oscillator the $\kappa $-deformed
energy-momentum dispersion relations satisfying the conditions
\begin{eqnarray}
\lim_{\kappa \rightarrow \infty }\omega _{\kappa }^{1}(\overrightarrow{p},%
\overrightarrow{q}) &=&\lim_{\kappa \rightarrow \infty }\widetilde{\omega }%
_{\kappa }^{1}(\overrightarrow{p},\overrightarrow{q})=\omega (%
\overrightarrow{p})=\sqrt{\overrightarrow{p}^{2}+m^{2}},  \label{f1} \\
\lim_{\kappa \rightarrow \infty }\omega _{\kappa }^{2}(\overrightarrow{p},%
\overrightarrow{q}) &=&\lim_{\kappa \rightarrow \infty }\widetilde{\omega }%
_{\kappa }^{2}(\overrightarrow{p},\overrightarrow{q})=\omega (%
\overrightarrow{q})=\sqrt{\overrightarrow{q}^{2}+m^{2}}.  \label{f2}
\end{eqnarray}%
Besides we have
\begin{eqnarray}
\lim_{\kappa \rightarrow \infty }\overrightarrow{P}(\overrightarrow{p},%
\overrightarrow{q}) &=&\lim_{\kappa \rightarrow \infty }\overrightarrow{S}(%
\overrightarrow{p},\overrightarrow{q})=\overrightarrow{p},  \label{p1} \\
\lim_{\kappa \rightarrow \infty }\overrightarrow{Q}(\overrightarrow{p},%
\overrightarrow{q}) &=&\lim_{\kappa \rightarrow \infty }\overrightarrow{R}(%
\overrightarrow{p},\overrightarrow{q})=\overrightarrow{q}.  \label{p2}
\end{eqnarray}

We assume that $\widehat{P}_{\mu }\triangleright a(p)=p_{\mu }a(p)$; from
the consistency of (\ref{ger}) with the $\kappa $-deformed addition law for
the fourmomenta (see(\ref{addd}), (\ref{amod})) one gets the following
constraints for the functions (\ref{fo1})-(\ref{fo2})
\begin{equation}
\overrightarrow{P}\mathrm{e}^{\omega _{\kappa }^{2}/2\kappa }+%
\overrightarrow{Q}\mathrm{e}^{-\omega _{\kappa }^{1}/2\kappa }=%
\overrightarrow{R}\mathrm{e}^{\widetilde{\omega }_{\kappa }^{1}/2\kappa }+%
\overrightarrow{S}\mathrm{e}^{-\widetilde{\omega }_{\kappa }^{2}/2\kappa },
\label{sp}
\end{equation}%
and
\begin{equation}
P_{0}+Q_{0}=R_{0}+S_{0},\qquad \Longleftrightarrow \qquad \omega _{\kappa
}^{1}+\omega _{\kappa }^{2}=\widetilde{\omega }_{\kappa }^{1}+\widetilde{%
\omega }_{\kappa }^{2}.  \label{ep}
\end{equation}

We shall now consider three classes of solutions of the eq. (\ref{sp})-(\ref%
{ep}) determining three types of $\kappa $-deformed oscillators.

\begin{enumerate}
\item \textit{The }$\kappa $\textit{-deformation of oscillator algebra} (\ref%
{ka1})-(\ref{kappaccr2}) \textit{obtained by the modification of
multiplication law} \cite{mpla},\cite{jpa}.
\end{enumerate}

We choose
\begin{eqnarray}
\overrightarrow{P}(\overrightarrow{p},\overrightarrow{q}) &=&\overrightarrow{%
p}\mathrm{e}^{-\omega _{\kappa }^{2}/2\kappa },\qquad \qquad \overrightarrow{%
Q}(\overrightarrow{p},\overrightarrow{q})=\overrightarrow{q}\mathrm{e}%
^{\omega _{\kappa }^{1}/2\kappa },  \label{va1} \\
\overrightarrow{S}(\overrightarrow{p},\overrightarrow{q}) &=&\overrightarrow{%
p}\mathrm{e}^{\widetilde{\omega }_{\kappa }^{2}/2\kappa },\qquad \qquad ~%
\overrightarrow{R}(\overrightarrow{p},\overrightarrow{q})=\overrightarrow{q}%
\mathrm{e}^{-\widetilde{\omega }_{\kappa }^{1}/2\kappa }.  \label{va2}
\end{eqnarray}%
The relation (\ref{sp}) leads to the identity
\begin{equation}
\overrightarrow{p}+\overrightarrow{q}=\overrightarrow{q}+\overrightarrow{p},
\end{equation}%
i.e. we obtain classical Abelian addition law for the three-momenta.

In order to satisfy the relation (\ref{ep}) we have to put the energies $%
\omega _{\kappa }^{1},\omega _{\kappa }^{2},\widetilde{\omega }_{\kappa
}^{1},\widetilde{\omega }_{\kappa }^{2}$ on the mass shells (\ref{mshell})
which are deformed in the following way
\begin{eqnarray}
&&C_{2}^{\kappa }(\overrightarrow{P}\mathrm{e}^{\omega _{\kappa
}^{2}/2\kappa },\omega _{\kappa }^{1})=m^{2},\qquad \qquad ~~C_{2}^{\kappa }(%
\overrightarrow{Q}\mathrm{e}^{-\omega _{\kappa }^{1}/2\kappa },\omega
_{\kappa }^{2})=m^{2},  \label{sh} \\
&&C_{2}^{\kappa }(\overrightarrow{S}\mathrm{e}^{-\widetilde{\omega }_{\kappa
}^{2}/2\kappa },\widetilde{\omega }_{\kappa }^{1})=m^{2},\qquad \qquad
C_{2}^{\kappa }(\overrightarrow{R}\mathrm{e}^{\widetilde{\omega }_{\kappa
}^{1}/2\kappa },\widetilde{\omega }_{\kappa }^{2})=m^{2}.  \label{msh}
\end{eqnarray}

The set of equations (\ref{sh})-(\ref{msh}) provides $\kappa $-deformed
energy-momentum dispersion relations, satisfying respectively two coupled
pairs of nonlinear equations
\begin{eqnarray}
\omega _{\kappa }^{1} &=&\epsilon \omega _{\kappa }(\overrightarrow{P}%
\mathrm{e}^{\omega _{\kappa }^{2}/2\kappa }),\qquad \qquad \omega _{\kappa
}^{2}=\widetilde{\epsilon }\omega _{\kappa }(\overrightarrow{Q}\mathrm{e}%
^{-\omega _{\kappa }^{1}/2\kappa }),  \label{om1} \\
\widetilde{\omega }_{\kappa }^{1} &=&\eta \omega _{\kappa }(\overrightarrow{S%
}\mathrm{e}^{-\widetilde{\omega }_{\kappa }^{2}/2\kappa }),\qquad ~~~~%
\widetilde{\omega }_{\kappa }^{2}=\widetilde{\eta }\omega _{\kappa }(%
\overrightarrow{R}\mathrm{e}^{\widetilde{\omega }_{\kappa }^{1}/2\kappa }),
\label{om2}
\end{eqnarray}%
where $(\epsilon ,\widetilde{\epsilon })=\pm 1,(\eta ,\widetilde{\eta })=\pm
1$. If we insert in (\ref{om1})-(\ref{om2}) the values (\ref{va1})-(\ref{va2}%
) we obtain in terms of variables $(\overrightarrow{p},\overrightarrow{q})$
the following expressions (see (\ref{mshell}))
\begin{equation}
\omega _{\kappa }^{1}=\widetilde{\omega }_{\kappa }^{1}=\omega _{\kappa }(%
\overrightarrow{p}),\qquad \qquad \omega _{\kappa }^{2}=\widetilde{\omega }%
_{\kappa }^{2}=\omega _{\kappa }(\overrightarrow{q}),  \label{2ener}
\end{equation}%
and it is evident that the energy conservation rule (\ref{ep}) is satisfied
identically.

The $\kappa $-deformed oscillators algebra can be written for the choice (%
\ref{va1})-(\ref{va2}) in the form identical to the classical relations (\ref%
{ka1})-(\ref{kappaccr2})
\begin{eqnarray}
&&[\;a_{\kappa }(p),a_{\kappa }(q)\;]_{\circ }=[\;a_{\kappa }^{\dag
}(p),a_{\kappa }^{\dag }(q)\;]_{\circ }=0\,,  \label{b1} \\
&&[\;a_{\kappa }^{\dag }(p),a_{\kappa }(q)\;]_{\circ }=\delta ^{(3)}(\vec{p}-%
\vec{q})\;,  \label{b2}
\end{eqnarray}%
where $[A.B]_{\circ }=A\circ B-B\circ A$\ \ and the new $\circ $%
-multiplication is defined as follows
\begin{eqnarray}
a_{\kappa }({p})\circ a_{\kappa }({q}) &=&a_{\kappa }\left( \mathrm{e}^{-%
\frac{\omega _{\kappa }(\overrightarrow{q})}{2\kappa }}\vec{p},\omega
_{\kappa }(\overrightarrow{p})\right) a_{\kappa }\left( \mathrm{e}^{\frac{%
\omega _{\kappa }(\overrightarrow{p})}{2\kappa }}\vec{q},\omega _{\kappa }(%
\overrightarrow{q})\right) \;,  \label{circc} \\
a_{\kappa }^{\dag }({p})\circ a_{\kappa }^{\dag }({q}) &=&a_{\kappa }^{\dag
}\left( \mathrm{e}^{\frac{\omega _{\kappa }(\overrightarrow{q})}{2\kappa }}%
\vec{p},\omega _{\kappa }(\overrightarrow{p})\right) a_{\kappa }^{\dag
}\left( \mathrm{e}^{-\frac{\omega _{\kappa }(\overrightarrow{p})}{2\kappa }}%
\vec{q},\omega _{\kappa }(\overrightarrow{q})\right) \;,  \label{multi1} \\
a_{\kappa }^{\dag }({p})\circ a_{\kappa }({q}) &=&a_{\kappa }^{\dag }\left(
\mathrm{e}^{-\frac{\omega _{\kappa }(\overrightarrow{q})}{2\kappa }}\vec{p}%
,\omega _{\kappa }(\overrightarrow{p})\right) a_{\kappa }\left( \mathrm{e}^{-%
\frac{\omega _{\kappa }(\overrightarrow{p})}{2\kappa }}\vec{q},\omega
_{\kappa }(\overrightarrow{q})\right) \;,  \label{multi2} \\
a_{\kappa }({p})\circ a_{\kappa }^{\dag }({q}) &=&a_{\kappa }\left( \mathrm{e%
}^{\frac{\omega _{\kappa }(\overrightarrow{q})}{2\kappa }}\vec{p},\omega
_{\kappa }(\overrightarrow{p})\right) a_{\kappa }^{\dag }\left( \mathrm{e}^{%
\frac{\omega _{\kappa }(\overrightarrow{p})}{2\kappa }}\vec{q},\omega
_{\kappa }(\overrightarrow{q})\right) \;.  \label{multi3}
\end{eqnarray}

Let us consider now the algebra $(A(a);\circ )$ of the creation oscillators
with modified $\circ $-multiplication. It can be shown that the relation (%
\ref{circc}) (for simplicity we consider here only creation operators) can
be extended to $n$-ary products as follows

\begin{equation}
a_{\kappa }({p}^{(1)})\circ _{\kappa }\cdots \circ _{\kappa }a_{\kappa }({p}%
^{(n)})=\prod_{k=1}^{n}a_{\kappa }\left( \chi _{n}^{(k)}(p_{0}^{(1)},\ldots
,p_{0}^{(n)})\vec{p}^{~(k)},p_{0}^{(k)}\right) ,  \label{mmmm}
\end{equation}%
where
\begin{equation}
\chi _{n}^{(k)}(p_{0}^{(1)},\ldots ,p_{0}^{(n)})=\exp \frac{1}{2\kappa }%
\left( {\sum\limits_{j=1}^{k-1}}p_{0}^{(j)}-{\sum\limits_{j=k+1}^{n}}%
p_{0}^{(j)}\right) \;.  \label{cchi}
\end{equation}%
From (\ref{mmmm}) one can show that the $n$-ary product is symmetric under
the change of order of oscillators, i.e. it satisfies the classical bosonic
relations
\begin{eqnarray}
&&a_{\kappa }(p_{1})\circ \cdots \circ a_{\kappa }(p_{i})\circ \cdots \circ
a_{\kappa }(p_{j})\circ \cdots \circ a_{\kappa }(p_{n})  \label{bozz} \\
&&\qquad \qquad =a_{\kappa }(p_{1})\circ \cdots \circ a_{\kappa
}(p_{j})\circ \cdots \circ a_{\kappa }(p_{i})\circ \cdots \circ a_{\kappa
}(p_{n})\;,  \notag
\end{eqnarray}%
and is associative, i.e. $\Phi =1$.

In particular for $n=3$ the associative triple product is the following
\begin{eqnarray}
&&a_{\kappa }(p)\circ a_{\kappa }({q})\circ a_{\kappa }({r})\equiv a_{\kappa
}(\overrightarrow{{p}}\mathrm{e}^{-\frac{1}{2\kappa }(\omega _{\kappa }(%
\overrightarrow{q})+\omega _{\kappa }(\overrightarrow{r}))},\omega _{\kappa
}(\overrightarrow{p}))a_{\kappa }(\overrightarrow{{q}}\mathrm{e}^{\frac{1}{%
2\kappa }(\omega _{\kappa }(\overrightarrow{p})-\omega _{\kappa }(%
\overrightarrow{r}))},\omega _{\kappa }(\overrightarrow{q}))  \notag \\
&&\qquad \qquad \qquad \qquad \qquad \qquad a_{\kappa }(\overrightarrow{{r}}%
\mathrm{e}^{\frac{1}{2\kappa }(\omega _{\kappa }(\overrightarrow{p})+\omega
_{\kappa }(\overrightarrow{q}))},\omega _{\kappa }(\overrightarrow{r})),
\end{eqnarray}%
and leads to the set of classical trilinear symmetry relations
\begin{equation}
a_{\kappa }({p}_{1})\circ a_{\kappa }({p}_{2})\circ a_{\kappa }({p}%
_{3})=a_{\kappa }({p}_{i})\circ a_{\kappa }({p}_{j})\circ a_{\kappa }({p}%
_{k}),
\end{equation}%
where $(i,j,k)$ is an arbitrary permutation of $(1,2,3).$

The relation (\ref{mmmm}) can be extended to the algebra $(A(a,a^{\dagger
});\circ )$ of arbitrary products of creation and annihilation operators
\cite{mpla},\cite{jpa} with the relations (\ref{bozz}) valid as well for
annihilation operators $a^{\dagger }(p)$. Due to the new $\kappa $-deformed
multiplication (\ref{circc}-\ref{multi3}) which leads to the\ symmetry
relations (\ref{bozz}), in the algebra $(A(a,a^{\dagger });\circ )$ the
braided structure is trivial. We add that using the relations (\ref{b1}-\ref%
{b2}) one can introduce also the normal products of $\kappa $-deformed field
oscillators.

\begin{description}
\item[2] \textit{The }$\kappa $\textit{-deformation introduced by nontrivial
involutive braid factor} \cite{am}, \cite{govv}.
\end{description}

Other choice of the functions occurring in (\ref{ger}) consistent with the
relation (\ref{sp}), is the following
\begin{eqnarray}
\overrightarrow{P}(\overrightarrow{p},\overrightarrow{q}) &=&\overrightarrow{%
p},\mathrm{\qquad \qquad \qquad }\overrightarrow{Q}(\overrightarrow{p},%
\overrightarrow{q})=\overrightarrow{q},  \label{2a1} \\
\overrightarrow{S}(\overrightarrow{p},\overrightarrow{q}) &=&\overrightarrow{%
p}\mathrm{e}^{\tilde{\omega}_{\kappa }^{2}/\kappa },\mathrm{\qquad \qquad }%
\overrightarrow{R}(\overrightarrow{p},\overrightarrow{q})=\overrightarrow{q}%
\mathrm{e}^{-\tilde{\omega}_{\kappa }^{1}/\kappa },  \label{2a2}
\end{eqnarray}%
where we shall show below that $\omega _{\kappa }^{1}=\widetilde{\omega }%
_{\kappa }^{1}=\omega _{\kappa }(\overrightarrow{p}),$ $\omega _{\kappa
}^{2}=\widetilde{\omega }_{\kappa }^{2}=\omega _{\kappa }(\overrightarrow{q}%
) $(see (\ref{mshell}) and (\ref{fo2})).

In order to obtain the conservation of energy (see (\ref{ep})) we postulate
that the relations (\ref{2ener}) are valid. We obtain the following
nonsymmetric addition law for the three-momenta (see also (\ref{addd})),
which as follows from (\ref{sp}) gives the same result after the exchange $%
1\leftrightarrow 2$ of two particles

\begin{equation}
\vec{p}^{(1+2)}\equiv \overrightarrow{p}\dotplus \overrightarrow{q}=\vec{p}%
\mathrm{e}^{\omega _{\kappa }(\overrightarrow{q})/2\kappa }+\overrightarrow{q%
}\mathrm{e}^{-\omega _{\kappa }(\overrightarrow{p})/2\kappa }=\vec{p}%
^{(2+1)}.  \label{add2}
\end{equation}

The binary oscillator algebra (\ref{ger}) for the choice (\ref{2a1})-(\ref%
{2a2}) of deformed momentum arguments takes the braided form (further we
denote $a(\overrightarrow{p})\equiv a(\overrightarrow{p},p_{0}=\omega
_{\kappa }(\overrightarrow{p})),$ $a^{\dagger }(\overrightarrow{p})\equiv a(%
\overrightarrow{p},p_{0}=-\omega _{\kappa }(\overrightarrow{p}))$

\begin{eqnarray}
a^{\dagger }(\overrightarrow{{p}})a(\overrightarrow{{q}})-\widehat{\tau }%
(a^{\dagger }(\overrightarrow{{p}})a(\overrightarrow{{q}})) &=&\delta ^{3}(%
\vec{p}\mathrm{e}^{\omega _{\kappa }(\overrightarrow{q})/2\kappa }-%
\overrightarrow{q}\mathrm{e}^{-\omega _{\kappa }(\overrightarrow{p})/2\kappa
})\equiv \delta ^{3}(\overrightarrow{p}\dot{-}\overrightarrow{q}),  \notag \\
a(\overrightarrow{{p}})a(\overrightarrow{{q}})-\widehat{\tau }(a(%
\overrightarrow{{p}})a(\overrightarrow{{q}})) &=&0,  \label{22a} \\
a^{\dagger }(\overrightarrow{{p}})a^{\dagger }(\overrightarrow{{q}})-%
\widehat{\tau }(a^{\dagger }(\overrightarrow{{p}})a^{\dagger }(%
\overrightarrow{{q}})) &=&0,  \notag
\end{eqnarray}%
where explicitly
\begin{eqnarray}
\widehat{\tau }[a(\overrightarrow{{p}})a(\overrightarrow{{q}})] &=&a(%
\overrightarrow{{q}}\mathrm{e}^{-\frac{1}{\kappa }\omega _{\kappa }(%
\overrightarrow{p})})a(\overrightarrow{{p}}\mathrm{e}^{\frac{1}{\kappa }%
\omega _{\kappa }(\overrightarrow{q})}),  \notag \\
\widehat{\tau }[a^{\dagger }(\overrightarrow{{p}})a^{\dagger }(%
\overrightarrow{{q}})] &=&a^{\dagger }(\overrightarrow{{q}}\mathrm{e}^{\frac{%
1}{\kappa }\omega _{\kappa }(\overrightarrow{p})})a^{\dagger }(%
\overrightarrow{{p}}\mathrm{e}^{-\frac{1}{\kappa }\omega _{\kappa }(%
\overrightarrow{q})}),  \label{22b} \\
\widehat{\tau }[a^{\dagger }(\overrightarrow{{p}})a(\overrightarrow{{q}})]
&=&a(\overrightarrow{{q}}\mathrm{e}^{\frac{1}{\kappa }\omega _{\kappa }(%
\overrightarrow{p})})a^{\dagger }(\overrightarrow{{p}}\mathrm{e}^{\frac{1}{%
\kappa }\omega _{\kappa }(\overrightarrow{q})}).  \notag
\end{eqnarray}%
The first relation (\ref{22a}) is the deformation of the relation (\ref%
{kappaccr2}), with the argument of Dirac delta deformed in agreement with (%
\ref{add2}). It is easy to see that from $\overrightarrow{p}-\overrightarrow{%
q}=0$ does not follow that $\overrightarrow{p}\dot{-}\overrightarrow{q}=0$,
as follows from nonsymmetric addition law (\ref{add2}) and replacement $%
\overrightarrow{q}$ $\rightarrow -\overrightarrow{q}$. The second relation
follows from (\ref{ger}), (\ref{2a1}),(\ref{2a2}) and the third relation is
its complex conjugation. It can be shown that the braid (\ref{22b}) is
involutive, i.e. $\widehat{\tau }^{2}=1$, as follows e.g. from the relation

\begin{equation}
\widehat{\tau }[a(\overrightarrow{{q}}\mathrm{e}^{-\frac{1}{\kappa }\omega
_{\kappa }(\overrightarrow{p})})a(\overrightarrow{{p}}\mathrm{e}^{\frac{1}{%
\kappa }\omega _{\kappa }(\overrightarrow{q})})]=a(\overrightarrow{{p}})a(%
\overrightarrow{{q}}).
\end{equation}

In order to obtain the relation (\ref{ep}) the energy-momentum dispersion
relations expressing the energies $\omega _{\kappa }^{1},\omega _{\kappa
}^{2},\widetilde{\omega }_{\kappa }^{1},\widetilde{\omega }_{\kappa }^{2}$
as the functions of three-momenta $\overrightarrow{P},\overrightarrow{Q},%
\overrightarrow{R},\overrightarrow{S}$ (see (\ref{ger})) should be
postulated as follows
\begin{eqnarray}
C_{2}^{\kappa }(\overrightarrow{P},\omega _{\kappa }^{1}) &=&m^{2},\qquad
\qquad C_{2}^{\kappa }(\overrightarrow{Q},\omega _{\kappa }^{2})=m^{2}, \\
C_{2}^{\kappa }(\overrightarrow{S}\mathrm{e}^{-\tilde{\omega}_{\kappa
}^{2}/\kappa },\widetilde{\omega }_{\kappa }^{1}) &=&m^{2},\qquad \qquad
C_{\kappa }^{2}(\overrightarrow{R}\mathrm{e}^{\tilde{\omega}_{\kappa
}^{1}/\kappa },\widetilde{\omega }_{\kappa }^{2})=m^{2}.
\end{eqnarray}%
After substitute of (\ref{2a1})-(\ref{2a2}) we obtain the confirmation that $%
\omega _{\kappa }^{1}=\widetilde{\omega }_{\kappa }^{1}=\omega _{\kappa }(%
\overrightarrow{p}),$ $\omega _{\kappa }^{2}=\widetilde{\omega }_{\kappa
}^{2}=\omega _{\kappa }(\overrightarrow{q})$.

In order to extend the relations (\ref{22a}) to arbitrary products of the
creation operators we should use the relations (\ref{brrr}) for the braid $%
\Psi \equiv \widehat{\tau }$. For ternary product we obtain for example the
following braiding relation
\begin{eqnarray}
a(\overrightarrow{{p}})a(\overrightarrow{{q}})a(\overrightarrow{{r}})
&\equiv &\widehat{\tau }_{12}[\widehat{\tau }_{23}(a(\overrightarrow{{p}})a(%
\overrightarrow{{q}})a(\overrightarrow{{r}})]  \label{rre3} \\
&=&a(\overrightarrow{{r}}\mathrm{e}^{-\frac{1}{\kappa }(\omega _{\kappa }(%
\overrightarrow{p})+\omega _{\kappa }(\overrightarrow{q}))})a(%
\overrightarrow{{p}}\mathrm{e}^{\frac{1}{\kappa }(\omega _{\kappa }(%
\overrightarrow{r})-\omega _{\kappa }(\overrightarrow{q}))})a(%
\overrightarrow{{q}}\mathrm{e}^{\frac{1}{\kappa }(\omega _{\kappa }(%
\overrightarrow{p})+\omega _{\kappa }(\overrightarrow{r}))}),  \notag
\end{eqnarray}%
which is consistent with the hexagonal relation.

In general case, if we permute the oscillators in $n$-ary product by
following the permutation $(1,2,\ldots ,n)\rightarrow (j_{1},j_{2},\ldots
j_{n})$, we should describe it as a product of simple braid factors
corresponding to the flips $(1,2,\ldots ,k,k+1,\ldots ,n)\rightarrow
(1,2,\ldots ,k+1,k,\ldots ,n).$ For\ every such simple flip we should
introduce the binary braiding factors $\widehat{\tau }_{kk+1}$\ given by the
relations (\ref{22b}). In such a case we obtain $n$-ary braiding as
expressed by suitable product of braiding factors.

\begin{description}
\item[3.] \textit{Braided }$\kappa $\textit{-deformed oscillators from }$%
\kappa $\textit{-deformed fourmomentum group composition law.}
\end{description}

Let us assume that in the relation (\ref{ger}) only one oscillator has
modified three-momentum dependence. We postulate
\begin{equation}
\overrightarrow{P}(\overrightarrow{p},\overrightarrow{q})=\overrightarrow{p},%
\mathrm{\qquad \qquad \qquad }\overrightarrow{Q}(\overrightarrow{p},%
\overrightarrow{q})=\overrightarrow{q}=\overrightarrow{R}(\overrightarrow{p},%
\overrightarrow{q}).  \label{3a1}
\end{equation}

It appears that the remaining three-momentum $\overrightarrow{S}$ can be
calculated from the conservation law (\ref{sp}). If we assume further the
relations (\ref{2ener}) one obtains
\begin{equation}
\overrightarrow{S}(\overrightarrow{p},\overrightarrow{q})=[\overrightarrow{p}%
\mathrm{e}^{\omega _{\kappa }(\overrightarrow{q})/2\kappa }-2\overrightarrow{%
q}\text{sinh}\frac{\omega _{\kappa }(\overrightarrow{p})}{2\kappa }]\mathrm{e%
}^{\omega _{\kappa }(\overrightarrow{q})/2\kappa },  \label{3a2}
\end{equation}%
which can be also obtained by the group manifold approach to the composition
of $\kappa $-deformed fourmomenta \cite{glik},\cite{glo}. We obtain the
following relation
\begin{equation}
a(\overrightarrow{{p}})a(\overrightarrow{{q}})\equiv \widehat{\rho }[a(%
\overrightarrow{{p}})a(\overrightarrow{{q}})]=a(\overrightarrow{{q}})a(%
\overrightarrow{S}(\overrightarrow{p},\overrightarrow{q})).  \label{3fac}
\end{equation}%
The operator $\widehat{\rho }$ describes a nonsymmetric braided factor,
which introduces only the $\kappa $-deformed momentum dependence in one
permuted oscillator. Further one can show that the braiding (\ref{3fac}) is
not involutive, i.e. $\widehat{\rho }^{2}\neq 1$.

The relation (\ref{3fac}) is consistent with the nonAbelian three-momentum
conservation law (\ref{add2}). If we introduce four functions $(\epsilon
,\eta =\pm 1)$
\begin{equation}
\overrightarrow{S}^{(\epsilon ,\eta )}(\overrightarrow{p},\overrightarrow{q}%
)=[\overrightarrow{p}\mathrm{e}^{\eta \omega _{\kappa }(\overrightarrow{q}%
)/2\kappa }-2\epsilon \overrightarrow{q} \text{sinh}\frac{\omega _{\kappa }(%
\overrightarrow{p})}{2\kappa }]\mathrm{e}^{\eta \omega _{\kappa }(%
\overrightarrow{q})/2\kappa },
\end{equation}%
where $\overrightarrow{S}=\overrightarrow{S}^{(+.+)}$, one can supplement
the relation (\ref{3fac})%
\begin{eqnarray}
a^{\dagger }(\overrightarrow{{p}})a^{\dagger }(\overrightarrow{{q}}) &\equiv
&\widehat{\rho }[a^{\dagger }(\overrightarrow{{p}})a^{\dagger }(%
\overrightarrow{{q}})]=a^{\dagger }(\overrightarrow{{q}})a^{\dagger }(%
\overrightarrow{S}^{(-,-)}(\overrightarrow{p},\overrightarrow{q})),
\label{3fac2} \\
a^{\dagger }(\overrightarrow{{p}})a(\overrightarrow{{q}}) &\equiv &\widehat{%
\rho }[a^{\dagger }(\overrightarrow{{p}})a(\overrightarrow{{q}})]=a^{\dagger
}(\overrightarrow{{q}})a^{\dagger }(\overrightarrow{S}^{(-,+)}(%
\overrightarrow{p},\overrightarrow{q})).  \label{3fac3}
\end{eqnarray}%
The basic field oscillator algebra relation (\ref{kappaccr2}) is $\kappa $%
-deformed as follows
\begin{equation}
a^{\dagger }(\overrightarrow{{p}})a(\overrightarrow{{q}})-\widehat{\rho }%
[a^{\dagger }(\overrightarrow{{p}})a(\overrightarrow{{q}})]=\delta ^{3}(\vec{%
p}\mathrm{e}^{\omega _{\kappa }(\overrightarrow{q})/2\kappa }-%
\overrightarrow{q}\mathrm{e}^{-\omega _{\kappa }(\overrightarrow{p})/2\kappa
})\equiv \delta ^{3}(\overrightarrow{p}\dot{-}\overrightarrow{q}).
\end{equation}

The triple product of $\kappa $-deformed oscillators which satisfies the
hexagon condition is characterized by the braided relation
\begin{eqnarray}
&&a(\overrightarrow{{p}})a(\overrightarrow{{q}})a(\overrightarrow{{r}}%
)\equiv \widehat{\rho }_{12}[\widehat{\rho }_{23}(a(\overrightarrow{{p}})a(%
\overrightarrow{{q}})a(\overrightarrow{{r}})] \\
&&\qquad \qquad \qquad \qquad \qquad =a(\overrightarrow{{p}})a(%
\overrightarrow{S}^{(+,+)}(\overrightarrow{p},\overrightarrow{r}))a(%
\overrightarrow{S}^{(+,+)}(\overrightarrow{q},\overrightarrow{r})).  \notag
\end{eqnarray}%
Similarly as in previous example (see (\ref{22a})-(\ref{22b}), (\ref{rre3}))
one gets the braiding relation for $n$-ary product of the $\kappa $%
-oscillators as product of binary braiding relations (\ref{3fac}), (\ref%
{3fac2}), (\ref{3fac3}). In addition contrary to our first two examples of $%
\kappa $-deformed oscillators, the noncommutative braid $\widehat{\rho }$
does not satisfy the relation characterizing the permutation group ($%
\widehat{\rho }_{i+1}\widehat{\rho }_{i}\widehat{\rho }_{i+1}=\widehat{\rho }%
_{i}\widehat{\rho }_{i+1}\widehat{\rho }_{i})$. This result is supplemented
with the relation $\widehat{\rho }^{2}\neq 1$. In order to get the
consistency with the relation (\ref{sp}), which describes the conservation
of energy under braiding, one should again keep the energy values given by
formula (\ref{2ener}).

\section{Outlook}

The proper choice of the $\kappa $-deformation of field oscillators algebra
should be covariant under the action of $\kappa $-Poincare algebra (see (\ref%
{q})). In particular if $\kappa $-deformed algebra is described by the
following braid relation
\begin{equation}
a(\overrightarrow{{p}})a(\overrightarrow{{q}})=\widehat{\tau }_{\kappa
}\vartriangleright \lbrack a(\overrightarrow{{p}})a(\overrightarrow{{q}})],
\label{ogdef}
\end{equation}%
it should have the same form in all $\kappa $-Poincare frames, i.e. ($%
\widehat{g}_{A}\equiv\widehat{ P}_{\mu },\widehat{M}_{\mu \nu }$)
\begin{equation}
\widehat{g}_{A}\vartriangleright \{\widehat{\tau }_{\kappa
}\vartriangleright \lbrack a(\overrightarrow{{p}})a(\overrightarrow{{q}})]\}=%
\widehat{\tau }_{\kappa }\vartriangleright \{\widehat{g}_{A}%
\vartriangleright \lbrack a(\overrightarrow{{p}})a(\overrightarrow{{q}})]\}.
\label{covq}
\end{equation}

The relation (\ref{sp}) provides the equality (\ref{covq}) for $\widehat{g}%
_{A}=\widehat{P}_{\mu }$, i.e. it represents the $\kappa $-deformed
translation invariance for the $\kappa $-deformed oscillators algebra. The
relation (\ref{covq}) for $\widehat{g}_{A}=\widehat{M}_{i}=\frac{1}{2}%
\epsilon _{ijk}\widehat{M}_{jk}$ is obvious. The explicit formula for $%
\widehat{\tau }_{\kappa }$ which satisfies (\ref{covq}) for the boost
generators $\widehat{N}_{i}=\widehat{M}_{i0}$, has been only calculated in
lowest orders of $\frac{1}{\kappa }$ \cite{yz1}.

The straightforward way which would provide the $\kappa $-covariant
deformation (\ref{ogdef}) is to construct the universal R-matrix for $\kappa
$-deformed Poincare algebra. In present literature such construction has
been achieved only in the category of triangular quasi-Hopf algebras $%
H(A,\cdot ,\Delta ,S,R,\Phi )$\ considered in detail by Drinfeld \cite{drin}%
, where the threelinear map $\Phi $ describes the coassociator modifying the
relations (\ref{trih})
\begin{equation}
(1\otimes \Delta )R=\Phi _{231}^{-1}R_{13}\Phi _{213}R_{12}\Phi
_{123}^{-1},\qquad (\Delta \otimes 1)R=\Phi _{321}R_{13}\Phi
_{132}^{-1}R_{23}\Phi _{123}.
\end{equation}

In \cite{yz4} there was calculated the coassociator $\Phi $ up to $\frac{1}{%
\kappa ^{3}}$ terms, and further shown that the universal $R$-matrix $R$ is
triangular, i.e. satisfies the condition $R_{21}=R^{-1}$. Because all
triangular algebras can be described by twist, one can conclude that the $%
\kappa $-Poincare algebra is a triangular quasi-Hopf algebra which can be
obtained by twist from undeformed Poincare algebra, but such a twist does
not satisfy the two-cocycle condition \cite{yz4},\cite{drin}, \cite{eme}.

In order to introduce the complete covariant $\kappa $-deformed oscillator
algebra we should define the binary algebra relation by introducing braided
factor expressed by the universal $R$-matrix, and incorporate the
coassociator $\Phi $ in the construction of algebraic multi-linear
relations. Binary twist factor and coassociator characterize the complete $%
\kappa $-deformed oscillator algebra \cite{yz4}. For achieving such a goal
in explicit form it is necessary the knowledge of universal R-matrix (see(%
\ref{bra})) and coassociator to arbitrary order in powers of $\frac{1}{%
\kappa }$.

\bigskip \textbf{Acknowledgements.} The authors acknowledge the support of
Polish Ministery of Science and Higher Education by grant NN202318534.

\end{document}